\documentclass[prl,twocolumn,superscriptaddress,showpacs]{revtex4}

\usepackage{psfig}

\begin{document}

\title{Noise-driven transition to quasi-deterministic limit cycle
dynamics in excitable systems}

\author{Cyrill B. Muratov}

\affiliation{Department of Mathematical Sciences, New Jersey Institute
of Technology, Newark, NJ 07102}

\author{Eric Vanden-Eijnden}

\affiliation{Courant Institute of Mathematical Sciences, New York
  University, New York, NY 10012}

\author{Weinan E}

\affiliation{Department of Mathematics and PACM, Princeton University,
Princeton, NJ 08544}

\date{\today}

\begin{abstract}
  
  The effect of small-amplitude noise on excitable systems with large
  time-scale separation is analyzed. It is found that small random
  perturbations of the fast excitatory variable result in the onset of
  a quasi-deterministic limit cycle behavior, absent without noise.
  The limit cycle is established at a critical value of the
  amplitude of the noise, and its period is nontrivially determined by
  the relationship between the noise amplitude and the time scale ratio.  It
  is argued that this effect might provide a mechanism by which the
  function of biological systems operating in noisy environments can
  be robustly controlled by the level of the noise.
\end{abstract}


\maketitle

Understanding the effect of random perturbations in dynamical systems
is a fundamental problem with a large number of applications in
physical, chemical, and biological sciences. When these perturbations
are small, noise-driven systems may often exhibit rare events which
lead to switching-like dynamics on long time-scales
\cite{freidlin}. This behavior is well understood in systems obeying
detailed balance, which is typical of systems close to thermal
equilibrium (see, for example, \cite{gardiner}).

Systems far from equilibrium often fail to obey detailed balance. Even
in the absence of noise, these systems are capable of having complex
dynamical behaviors. One example of such a behavior is limit cycle
oscillations. Another example is excitability: in excitable systems,
small perturbations of the unique stable steady state decay, while
sufficiently large perturbations lead to large-amplitude dynamical
excursions before the system returns to the steady state (see, for
example, \cite{mikhailov}). Excitable systems are very common in
biology, with nerve cells being just one example (see, for example,
\cite{keener}).

The purpose of this Letter is to show that under certain conditions
small noise may transform the dynamics of an excitable system into a
{\em quasi-deterministic limit cycle}, signifying a transition from
excitable to oscillatory behavior under the action of the noise. This
transition occurs at a critical value of the noise amplitude.
Furthermore, the frequency and other parameters of the limit cycle are
controlled by the amplitude of the noise, which distinguishes this
mechanism from other mechanisms of noise-induced coherence
(see, e.g., \cite{gang93,pikovsky97,osipov00}). 


The key to our argument is the existence of a strong time scale
separation in the system's dynamics. In biology, this is typically the
case: excitability often arises as a result of the competition of
positive and negative feedbacks operating on different time-scales,
with fast excitatory and slow recovery variables \cite{keener}. In
this situation it is possible to have an interplay between the slow
deterministic time-scale of the recovery variable and the
exponentially long time-scale associated with rare events induced by
the noise. It is precisely this interplay that makes the considered
effect possible.








Consider a general dynamical system
\begin{eqnarray}
  \dot u & = & f(u, v) + \sqrt{\varepsilon}\, \eta, \label{u} \\ 
  \dot v & = & \alpha g(u, v). \label{v}
\end{eqnarray}
In the context of excitable systems, $u$ and $v$ (generally, vectors)
will be sets of excitatory and recovery variables, respectively; $f$
and $g$ are the nonlinearities, $\alpha$ is the ratio of the
time-scales, $\eta$ is some external noise perturbing the excitatory
variables, and $\varepsilon\ll 1$ measures its amplitude. If $\alpha
\ll 1$, there is a large time-scale separation in the deterministic
part of the dynamics governed by $f$ and $g$, with $u$ and $v$ being
the ``fast'' and the ``slow'' variables, respectively \footnote{This
  distinction is only legitimate when both $u$ and $v$ are $O(1)$.
  During a large excursion, it may no longer remain the case.}. That
is, on the time-scale of order 1 the dynamics of $u$ is governed by
(\ref{u}) with $v$ frozen. Therefore, in the absence of the noise, the
trajectory quickly approaches the slow manifold defined by the
solutions of $f(u, v) = 0$ for fixed $v$, and then proceeds on the
time scale $\alpha^{-1}$ along this slow manifold and into the
globally attracting equilibrium point.

With the introduction of the noise, the situation changes. Indeed, the
trajectory may escape the slow manifold in the $u$ direction via a
noise-activated process on the Arrhenius time-scale $\tau = \nu^{-1}
\exp ( \beta(v)/ \varepsilon) \gg 1$, where $\beta(v)$ is some energy
barrier to be crossed to initiate the escape and $\nu$ is some
characteristic frequency \cite{freidlin}. As a result, the system can
perform a large excursion driven by the deterministic part of the
dynamics, after which the trajectory can land again somewhere on the
slow manifold.  The dynamics can then proceed along the slow manifold
until another escape event happens, and so on. This process may lead
to a \textit{bona-fide} limit cycle because of the following two
ingredients. First, the interplay between the escape events and the
motion along the slow manifold requires that their time-scales be
comparable. But since the escape rate is a rapidly varying function of
$v$, escape from the slow manifold will occur with overwhelming
probability in the small vicinity of point $v = v_\star$ on this
manifold, where $v_\star$ satisfies
\begin{eqnarray} 
  \label{beta}
 \beta(v_\star) = \varepsilon \log \alpha^{-1}.
\end{eqnarray}
Here we assumed that $\alpha,\varepsilon \to 0$ in a way that
$\varepsilon \log \alpha^{-1}$ remains finite and neglected higher
order terms.  The next ingredient to obtain a limit cycle behavior is
then a mechanism of reset. It requires that after a large excursion
initiated from $v_\star$, the trajectory returns to the slow manifold
to a point which leads again to $v_\star$ by the slow motion. Then the
process will repeat itself indefinitely and the dynamics of the system
will indeed be a limit cycle.  Let us point out that the mechanism of
escape at a fixed $v_\star$ is related to the phenomenon of stochastic
resonance (for a review, see \cite{gammaitoni98}). 
Also note that the possibility of limit cycle behavior under the
action of small noise in systems lacking such behavior in the absence
of the noise was recently pointed out by Freidlin
\cite{freidlin01}. 

\begin{figure}
\centerline{\psfig{figure=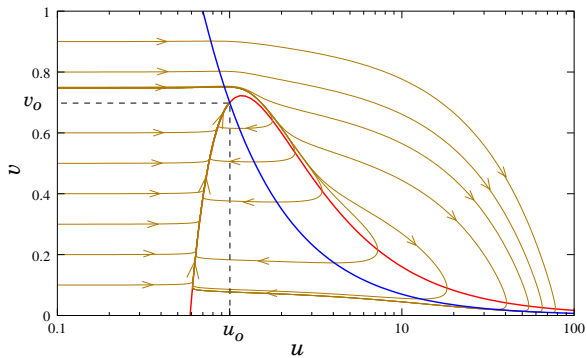,width=3.5in}}
\caption{Deterministic flow generated by the nonlinearities from
Eq.~(\ref{brus}). Results of the numerical solution of Eqs.~(\ref{u})
and (\ref{v}) in the $u-v$ plane for $A = 0.7$ and $\alpha = 0.01.$
Thick red and blue curves are the $u$- and $v$-nullclines,
respectively.}
\label{f:phase}
\end{figure}

To demonstrate the feasibility of the scenario above, we will consider
the Brusselator \cite{nicolis}. This is a prototypical model of a
system far from equilibrium in which the nonlinearities take on the
following form
\begin{eqnarray}
  f = 1 + u^2 v - (1 + A) u, \quad g = A u - u^2 v. \label{brus}
\end{eqnarray}
Here $u$ and $v$ are scalars and $A$ is a control parameter. For
simplicity we will assume that $\eta$ is external white Gaussian
noise, with $\langle \eta(t) \eta(t') \rangle = \delta(t - t')$.


The Brusselator is an excitable system when $\alpha \ll 1$ and $A <
1$. This can be seen from its phase portrait shown in
Fig.~\ref{f:phase}. When $A < 1$, the nullclines of Eqs.~(\ref{u}) and
(\ref{v}) intersect on the stable branch of the $u$-nullcline, so the
flow is always into the unique equilibrium point
\begin{eqnarray}
  u_0 = 1, \qquad v_0 = A.
\end{eqnarray}
Note that the slow manifold here is the rising part of the
$u$-nullcline (see Fig.~\ref{f:phase}). It is also clear from the
figure that sufficiently large increases in the $u$ variable away from
equilibrium will result in large excursions.


\begin{figure*}
  \centerline{\psfig{figure=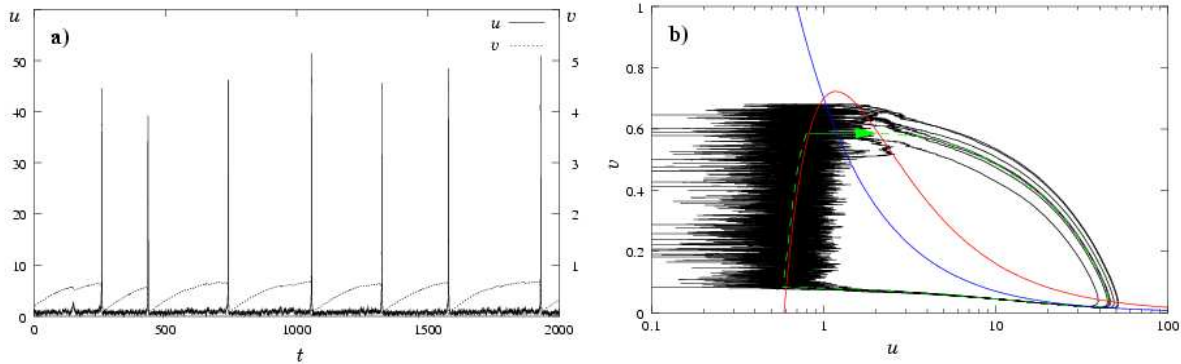,width=6.5in}}
  \caption{Numerical solution of the stochastic differential
  equation. The parameters are: $A = 0.7, \alpha = 0.01, \varepsilon =
  0.1$. (a) The time series with the quasi-periodic spike train. (b)
  The phase plane plot showing the quasi-deterministic limit
  cycle. Red and blue lines are the $u$- and $v$-nullclines,
  respectively. Solid green line is the predicted escape path and
  dashed green line is the excursion following the escape. }
  \label{f:stoch}
\end{figure*}

We performed numerical simulations of the stochastic differential
equations given by Eqs.~(\ref{u}), (\ref{v}), and (\ref{brus}).
Fig.~\ref{f:stoch}(a) shows the time series for one realization of the
noise. This figure shows a train of large amplitude spikes in the
excitatory variable. What is striking is that the spikes are occurring
in an almost periodic fashion, with their amplitude and other
characteristics being approximately the same. This can be better seen
from the phase plane, Fig.~\ref{f:stoch}(b). In fact,
Fig.~\ref{f:stoch}(b) strongly suggests a nearly limit cycle behavior
in the $u$ and $v$ variables. To verify this, we computed the
stationary probability distribution function (PDF) for this process
(not shown).
The PDF is found to be concentrated around a part of the slow
manifold. One can also see a pronounced loop in the PDF for large
values of $u$. Importantly, the PDF has a maximum at some value of $v
= v_\star < v_0$. Let us point out that this result shows that our
system lies beyond the region of applicability of the large deviations
theory (if the latter were true, the PDF would be monotonically
decreasing away from the equilibrium point \cite{freidlin}).

\begin{figure}[b]
\centerline{\psfig{figure=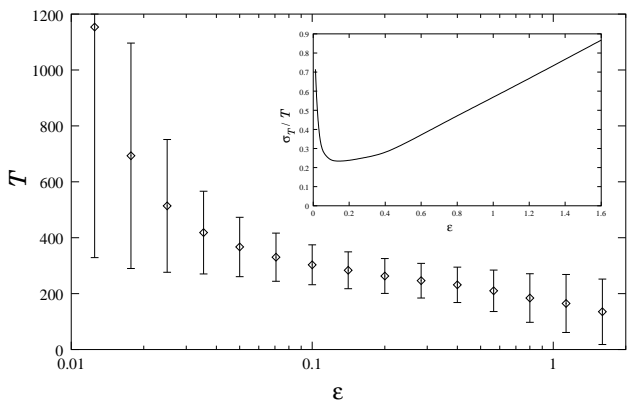,width=3.25in}}
\caption{Mean interspike distance $T$ and its standard deviation
$\sigma_T$ (shown as errorbars) as a function of the noise
amplitude. Inset: the plot of the ratio of the standard deviation to
the mean. Except for the noise, the parameters are as in
Fig. \ref{f:stoch}.}
\label{f:inter}
\end{figure}

To further investigate the effect of the noise, we have analyzed the
statistics of the interspike time intervals for different values of
the noise amplitude $\varepsilon$. For the purpose of this analysis,
we defined as a spike any excursion with an amplitude $u_\mathrm{max}
\geq 10$. Fig.~\ref{f:inter} shows the mean interspike distance $T$
and its standard deviation $\sigma_T$ obtained from the numerical
solution of Eqs.~(\ref{u}), (\ref{v}) and (\ref{brus}) for different
value of the amplitude of the noise. Also, in the inset we show the
ratio $\sigma_T / T$ as a function of $\varepsilon$, which
characterizes the ``signal-to-noise ratio'' for the interspike
distance.

First, observe that for very small $\varepsilon$ the spikes are
exceedingly rare and have the character of a Poisson process (see the
inset). They represent rare large-amplitude fluctuations away from the
equilibrium point.  Then, at some critical value of $\varepsilon
\simeq 0.02$ the increase in the noise amplitude results in a rapid
decrease in the interspike distance.  Furthermore, the ratio $\sigma_T
/ T$ rapidly decreases and stays low in a broad range $0.05 \lesssim
\varepsilon \lesssim 0.5$, signifying high degree of signal coherence
(see Fig.~\ref{f:stoch}). Let us emphasize that in this range $T$
shows significant dependence on $\varepsilon$, while $\sigma_T$ does
not (as seen from the errorbars). In other words, the noise amplitude
really acts as a control parameter for the deterministic oscillatory
behavior of the system. Finally, for larger noise amplitude, the spike
train looses coherence again, since then the noise is no longer weak.

To explain these observations and corroborate the general scenario
given above, let us consider the limit of strong time-scale
separation. On the fast time-scale the value of $v$ is fixed, hence
Eq.~(\ref{u}) describes the motion of a particle in the potential well
of the form $V(u) = -\frac{1}{3} v u^3 + \frac{1}{2} (1 + A) u^2 - u$.
This is the classical escape problem of Kramers, for which the average
escape time $\tau$ is given by the following formula \cite{gardiner}
\begin{eqnarray}
  \tau = {2 \pi \over \sqrt{(1 + A)^2 - 4 v}} \exp \left\{ { [(1 + A)^2 - 4
      v]^{3/2} \over 3 v^2 \varepsilon} \right\}.
\end{eqnarray}
%
 From this equation, we obtain explicitly $\beta(v) = {[(1 + A)^2 - 4
v]^{3/2} / 3 v^2}$. This $\beta(v)$ is inserted into Eq.~(\ref{beta}),
which fixes $v_\star$ as a function of $\alpha$ and $\varepsilon$.

After the trajectory escapes the neighborhood of the stable nullcline
at $v = v_\star$, it continues moving toward increasing values of $u$.
At this point the effect of the noise becomes negligible. With the
increase of $u$, the effective time-scale of $v$ decreases [see
Eq.~(\ref{v})], so when the system undergoes a large excursion, the
time-scales of these two variables can no longer be separated. On the
other hand, when $u \gg 1$ and $v \sim 1$, Eqs.~(\ref{u}) and
(\ref{v}) can be simplified by neglecting all the terms except $u^2
v$. The resulting system of equations with the asymptotic boundary
conditions $u(-\infty) = 0$, $v(-\infty) = v_\star$ can be integrated
to give $v = v_\star - \alpha u$ and $\dot u = u^2 (v_\star - \alpha
u)$.  This equation can be solved exactly to give the transition layer
for the rising part of the spike. It shows that in the spike $u$ rises
to $u_\mathrm{max} = \alpha^{-1} v_\star \gg 1$ on the time-scale of
$\alpha \ll 1$, while $v$ approaches zero asymptotically. This is then
followed by the fall of the trajectory onto the $u$-nullcline with
fixed (small) $v$ on the slower time scale of order 1, see
Eq.~(\ref{u}) with $v = 0$ (a more precise layer analysis of the
solution can be performed \cite{lmo}).

Following the excursion, on the slow time-scale the trajectory moves
along the $u$-nullcline, starting at $v = 0$ (asymptotically). After a
little algebra we obtain for the slow motion:
\begin{eqnarray} \label{slow}
  \dot v = {2 \alpha (A - v) \over 1 + A - 2 v + \sqrt{(1 + A)^2 - 4
v}}.
\end{eqnarray}
 From Eq.~(\ref{slow}), $\dot v > 0$ for $0 < v < v_0$, so the
trajectory will reach the point $v = v_\star$ on the $u$-nullcline in
time $T = T(v_\star)$, which can be explicitly calculated by
integrating Eq.~(\ref{slow}). Upon reaching $v_\star$, the trajectory
jumps off from the nullcline and returns onto the $u$-nullcline at the
point where $v = 0$, thus completing the loop
[cf. Fig. \ref{f:stoch}(b)]. In summary, we obtain a limit cycle whose
amplitude and period are asymptotically determined by the value of
$v_\star$, which is in turn a function of $\alpha$ and $\varepsilon$.

Since the system spends most of the time on the slow manifold,
asymptotically the period of the limit cycle will be equal to
$T(v_\star)$. This prediction can be compared with the mean interspike
distance obtained from the simulations of Fig. \ref{f:inter}. We found
that the two are in qualitative agreement for $0.05 \lesssim \varepsilon
\lesssim 0.5$, but $T(v_\star)$ consistently underestimates the observed
value of $T$ by a factor of about 1.5. This is not unexpected, since
the small parameter of the asymptotics is $1/\ln \alpha^{-1}$, which
in practice is not very small. We verified that the accuracy increases
with decrease of $\alpha$.

The above analysis also predicts the existence of a critical amplitude
of the noise for the establishment of the limit cycle behavior.
Indeed, the attainable values of $v$ on the $u$-nullcline lie in the
interval $0 < v < v_0$. Therefore, the solution of Eq.~(\ref{beta})
may lie in this interval only if $\varepsilon > \varepsilon_c$, where
\begin{eqnarray}
  \varepsilon_c = {(1 - A)^3 \over 3 A^2 \log \alpha^{-1}}.
\end{eqnarray}
No limit cycle behavior is possible when $\varepsilon <
\varepsilon_c$. As $\varepsilon$ approaches $\varepsilon_c$ from
above, we have $v_\star \to v_0$, and $T(v_\star) \to
\infty$. Therefore, for fixed deterministic part of the dynamics there
is a transition to a limit cycle behavior at a critical value of the
amplitude of the noise. We verified that the predicted value of
$\varepsilon_c$ gives the correct order of magnitude for the onset of
the oscillatory behavior. 

Let us emphasize that the mechanism described above is robust: it does
not require fine-tuning of the system's parameters. This is in
contrast with other mechanisms, by which noise can induce coherence
(see, for example, \cite{gang93,pikovsky97,osipov00}). Those
mechanisms require that in the absence of the noise the system is near
the threshold between excitable and oscillatory behavior.

Now we discuss the potential implications of the observed phenomena.
First of all, our results suggest that in systems with strong
time-scale separation one should be careful in identifying the class
of dynamical models for explaining the observations of oscillatory
behaviors. Real systems are always noisy, and our analysis indicates
that oscillatory behavior can be obtained in systems with
intrinsically \textit{non-oscillatory} dynamics. In particular, this
may be relevant to the identification of cell types in model neural
networks (see for example the discussions in \cite{terman97,vilar02}).
We note that the noise-induced transition similar to the one discussed
in this Letter was recently observed in the numerical simulations of
randomly forced Hodgkin-Huxley neuron \cite{takahata02}.

Another possible implication has to do with coupled excitable
systems. Our analysis indicates that in such systems the level of the
noise, both extrinsic and intrinsic, may be used as an information
carrier and transformed into (quasi-)deterministic signal. As a
prototype, consider a system of all-to-all positively coupled
excitable cells. Under the action of the noise of sufficiently small
amplitude each cell will occasionally generate a spike. These spikes
will have random phases, so their total input on each individual cell
may average to a stationary random signal of low intensity. Now, if
the noise level suddenly increases due to an external disturbance, the
cells may switch to the noise-assisted oscillatory mode. This will
further increase the effective noise amplitude, so that the
oscillatory mode may persist even after the disturbance is removed. In
colloquial terms, the system in a dormant state may wake up from
the outside rattle.

In a similar way, our results may be applied to spatially distributed
excitable media \cite{mikhailov}. In these systems the analogue of the
noise-activated event will be the formation of the radially-symmetric
nucleus, leading to the consequent initiation of the
radially-divergent wave \cite{om:prl95,mo:epjb01}. In the wake of such
a wave the system will undergo recovery. It is clear, then, that the
system will be most recovered at the position where the wave was
initiated. Hence, the new wave will be initiated again at the same
spot, with the dynamics repeating periodically. This suggests that the
well-known phenomenon of target pattern formation in two-dimensional
excitable media \cite{mikhailov} might have an alternative
interpretation in terms of noise-driven quasi-periodic wave
generation.


C. B. M. is partially supported by NSF via grant
DMS02-11864. E. V.-E. is partially supported by NSF via grants
DMS01-01439, DMS02-09959 and DMS02-39625. W. E is partially supported
by NSF via grant DMS01-30107. 





\bibliography{../stat,../nonlin,../mura,../bio}

\end{document}